\documentclass[10pt, conference, compsocconf]{IEEEtran}

\usepackage{url}
\usepackage{verbatim}
\usepackage[noadjust]{cite}

\usepackage{graphics,graphicx}
\usepackage{amsthm,amsmath,amssymb,enumerate}
\usepackage{enumerate}
\usepackage{algorithm}
\usepackage{algorithmic}
\usepackage{array}
\newtheoremstyle{mystyle}
  {}
  {}
  {}
  {\parindent}
  {\textnormal}
  {.}
  { }
  {}

\theoremstyle{mystyle}
\newtheorem{mydef}{DEFINITION}
\newtheorem* {myprobstmt*}{PROBLEM STATEMENT}

\begin{document}

\title{Subgraph Similarity Search in Large Graphs}

%
\author{
    \IEEEauthorblockN{Kanigalpula Samanvi\IEEEauthorrefmark{1}, Naveen Sivadasan\IEEEauthorrefmark{2}}\\
    \IEEEauthorblockA{\IEEEauthorrefmark{1}Dept. of Computer Science and Engineering\\
   Indian Institute of Technology Hyderabad, India
    \\cs13m1001@iith.ac.in}\\
    \IEEEauthorblockA{\IEEEauthorrefmark{2}TCS Innovation Labs Hyderabad, India\\
    naveen@atc.tcs.com}
}

\maketitle

\begin{abstract}
One of the major challenges in applications related to social networks, computational biology, collaboration networks etc., is to efficiently search for similar patterns in their underlying graphs. These graphs are typically noisy and contain thousands of vertices and millions of edges. In many cases, the graphs are unlabeled and the notion of similarity is also not well defined.
We study the problem of searching an induced subgraph in a large target graph that is most similar to the given query graph. We assume that the query graph and target graph are undirected and unlabeled.  We use graphlet kernels \cite{shervashidze2009efficient} to define graph similarity. Graphlet kernels are known to perform better than other kernels in different applications.

 Our algorithm maps topological neighborhood information of vertices in the query and target graphs to vectors. These local topological informations are then combined to find a target subgraph having highly similar global topology with the given query graph. We tested our algorithm on several real world networks such as 
facebook network, google plus network, youtube network, amazon network etc.  Most of them contain thousands of vertices and million edges. 
Our algorithm is able to detect highly similar matches when queried in these networks. Our multi-threaded implementation takes about one second to find the match on a 32 core machine, excluding the time for one time preprocessing.
Computationally expensive parts of our algorithm can be further scaled to standard parallel and distributed frameworks like map-reduce.
\end{abstract}


\IEEEpeerreviewmaketitle

\section{Introduction}
Similarity based graph searching has attracted considerable attention in the context of social networks, road networks, collaboration networks, software testing, computational biology, molecular chemistry etc. In these domains, underlying graphs are large with tens of thousands of vertices and millions of edges. Subgraph searching is fundamental to the applications, where occurrence of the query graph in the large target graph has to be identified. Searching for exact occurrence of an induced subgraph isomorphic to the query graph is known as the subgraph isomorphism problem, which  is known to be NP-complete for undirected unlabeled graphs.

Presence of noise in the underlying graphs and need for searching `similar' subgraph patterns are characteristic to these applications. For instance, in computational biology, the data is noisy due to possible errors in data collection and different thresholds for experiments. In object-oriented programming, querying typical object usage patterns against the target object dependency graph of a program run can identify deviating locations indicating potential bugs \cite{nguyen2009graph}. In molecular chemistry, identifying similar molecular structures is a fundamental problem. Searching for similar subgraphs plays a crucial role in mining and analysis of social networks.
Subgraph similarity searching is therefore more natural in these settings in contrast to exact search. In subgraph similarity search problem, induced subgraph of the target graph that is `most similar' to the query graph has to be identified, where similarity is defined using some distance function.  
Quality of the solution and computational efficiency are two  major challenges in these search problems. 
In this work, we assume that both the underlying graph and query graph are unlabeled and undirected. 

Most applications work with a distance metric to define similarity between two entities (graphs in our case). Popular distance metrics include 
Euclidean distance, Hamming distance, Edit distance, Kernel functions  \cite{haussler1999convolution,desobry2005class,
kondor2003kernel,vishwanathan2004fast}
etc.
We use graph kernel functions to define graph similarity.

Kernels are symmetric functions that map pairs of entities from a domain to real values which indicate their similarity.
Kernels that are positive definite not only define similarity between pairs of entities but also allow implicit mapping of objects to a high-dimensional feature space and operating on this space without requiring to compute explicit mapping of objects in the feature space. Kernels implicitly yield inner products between the feature vectors without explicit computation of the same in feature space. This is usually computationally cheaper than explicit computation. This approach is usually referred to as the kernel trick or kernel method. 
Kernel methods have been widely applied to sequence data, graphs, text, images, videos etc., as many of the standard machine learning algorithms including support vector machine (SVM) and principle component analysis (PCA)  can directly work with kernels.


Kernels have been successfully applied in the past in the context of graphs \cite{hido2009linear,hammond2011wavelets,shervashidze2009fast}. There are several existing graph kernels based on various graph properties,  such as random walks in the graphs \cite{gartner2003graph,kashima2002kernels}, cyclic patterns  \cite{horvath2004cyclic}, graph edit distance \cite{neuhaus2005edit}, shortest paths \cite{borgwardt2005shortest,bunescu2005shortest}, frequency of occurrences of special subgraphs \cite{frohlich2005optimal,ramon2003expressivity,menchetti2005weighted} and so on. 

Graphlet kernels are defined based on occurrence frequencies of small induced subgraphs called graphlets in the given graphs \cite{shervashidze2009efficient}. Graphlet kernels have been shown to provide good SVM classification accuracy in comparison to random walk kernel and shortest path kernel on different data sets including protein and enzyme data \cite{shervashidze2009efficient}.  
Graphlet kernels are also of theoretical interest. It is known that under certain restricted settings, if two graphs have distance zero with respect to their graphlet kernel value then they are isomorphic \cite{shervashidze2009efficient}. 
Improving the efficiency of computing graphlet kernel is also studied in \cite{shervashidze2009efficient}.  Graphlet kernel computation can also be scaled to parallel and distributed setting in a fairly straight forward manner.
In our work, we use graphlet kernels to define graph similarity. 


\subsection{Related Work}

Similarity based graph searching has been studied in the past under various settings. In many of the previous works, it is  assumed that the graphs are labeled.
In one class of problems, a large database of graphs is given and the goal is to find the most similar match in the database with respect to the given query graph \cite{shasha2002algorithmics,yan2004graph,zhang2009gaddi,mongiovi2010sigma,zhang2010sapper,wang2009g}.
In the second class, given a target graph and a query graph, subgraph of the target graph that is most similar to the query graph needs to be identified \cite{khan2011neighborhood,khan2013neighborhood,sun2012efficient,yuan2015}. Different notions of similarity were also explored in the past for these classes of problems.

In \cite{tian2008tale}, approximate matching of query graph in a database of graphs is studied. The graphs are assumed to be labeled. Structural information of the graph is stored in a hybrid index structure based on B-tree index. Important vertices of a query graph are matched first and then the match is extended progressively. 
In \cite{zheng2013graph}, graph similarity search on labeled graphs from a large database of graphs under minimum edit distance is studied. 
In \cite{khan2011neighborhood}, algorithm for computing top-$k$ approximate subgraph matches for a given query graph in a large labeled target graph is given. In this work, the target graph is converted into a set of multidimensional vectors based on the labels in the vertex neighborhoods. Only matches above a user defined threshold are computed. With higher threshold values, the match is a trivial vertex to vertex label matching. In \cite{khan2013neighborhood}, label matching is performed while simultaneously preserving pairwise vertex proximity. Their query time is proportional to the product of number of vertices of the query and target graph.  Subgraph matching in a large target graph for graphs deployed on a distributed memory store was studied in \cite{sun2012efficient}. In \cite{yuan2015}, efficient distributed subgraph similarity search to retrieve matches whose number of missing edges is   below a given threshold is studied. It looks for exact matching and not similarity matching. Though different techniques were studied in the past for the problem of similarity searching in various settings, to the best of our knowledge, little work has been done on subgraph similarity search on large unlabeled graphs. In many of the previous works, either the vertices are assumed to be labeled or the graphs  they work with are small with hundreds of vertices.

\subsection{Our Contribution}
We consider undirected graphs with no vertex or edge labels. We use graphlet kernel to define similarity between graphs. We give a subgraph similarity matching algorithm that takes as input a large target graph and a query graph and identifies an induced subgraph of the target graph that is most similar to the query graph with respect to the graphlet kernel value.  

In our algorithm, we first compute vertex labels for vertices in both query and target graph. These labels are vectors in some fixed dimension and are computed based on local neighborhood structure of vertices in the graph. Since our vertex labels are vectors, unlike many of the other labeling techniques, our labeling allows us to define the notion of similarity between vertex labels of two vertices to capture the topological similarity of their corresponding neighborhoods in the graph.  We build a nearest neighbor data structure for vertices of the target graph based on their vertex labels. Computing vertex label for target graph vertices and building the nearest neighbor data structure are done in the preprocessing phase. Using nearest neighbor queries on this data structure, vertices of the target graph that are most similar to the vertices of the query graph are identified. Using this smaller set of candidate vertices of target graph, a seed match is computed for the query graph. Using this seed match as the basis, our algorithm computes the final match for the full query graph.

We study the performance of our algorithm on several real life data sets including facebook network, google plus network, 
youtube network, road network, amazon network provided by the Stanford Large Network Dataset Collection (SNAP) \cite{snapnets} and DBLP network \cite{DBLP}. We conduct number of experimental studies to measure the search quality  and run time  efficiency.
For instance, while searching these networks with their communities as query graphs, the computed match and the query graph  has similarity score close to 1, where 1 is the maximum possible similarity score. In about 30\% of the cases, our algorithm is able to identify the exact match and in about 80\% of the cases, vertices of exact match are present in the pruned set computed by the algorithm. We validate our results by showing that similarity scores between random subgraphs and similarity scores between random communities in these networks are significantly lower. We also query communities across networks and in noisy networks and obtain matches with significantly high similarity scores. We use our algorithm to search for dense subgraphs and identify subgraphs with significantly high density. 

Computationally expensive parts of our algorithm can be easily scaled to standard parallel and distributed computing frameworks such as map-reduce. 
Most of the networks in our experiments have millions of edges and thousands of vertices.
Our multithreaded implementation of the search algorithm takes close to one second on these networks on a 32 core machine for the search phase. This  excludes time taken by the one time pre-processing phase.



\section{Preliminaries}
Graph is an ordered pair $G=(V,E)$ comprising a set $V$ of vertices and a set $E$ of edges. To avoid ambiguity, we also use $V(G)$ and $E(G)$ to denote the vertex and edge set. We consider only undirected graphs with no vertex or edge labels.
A subgraph $H$ of $G$ is a graph whose vertices are a subset of $V$, and whose edges are a subset of $E$ and is denoted as $H \subseteq G$. 
An induced subgraph $G'$ is a graph whose vertex set $V'$ is a subset of $V$ and whose edge set is the set of all edges present in $G$ between vertices in $V'$.

\begin{mydef}[Graph Isomorphism]
Graphs $G_{1}$ and $G_{2}$ are isomorphic if there exists a bijection $b:V(G_{1}) \rightarrow V(G_{2}) $ such that any two vertices $u$ and $v$ of $G_{1}$ are adjacent in $G_{1}$ if and only if $b(u)$ and $b(v)$ are adjacent in $G_{2}$.
\end{mydef}  

\begin{mydef}[Subgraph Isomorphism]
Graph $G_1$ is isomorphic to a subgraph of graph $G_{2}$, if there is an induced subgraph of $G_2$ that is isomorphic to $G_1$.
\end{mydef}

\begin{mydef}[Graph Similarity Searching]
Given a collection of graphs and a query graph,  find graphs in the collection that are closest to the query graph with respect to a given distance function between graphs.
\end{mydef}

\begin{mydef}[Subgraph Similarity Searching]
Given graphs $G_{1}$ and $G_{2}$, determine a subgraph $G^* \subseteq G_{1}$ that is closest to $G_{2}$ with respect to a given distance function between graphs.
\end{mydef}

\subsection{Graphlet Kernel}
\begin{figure}[ht]
\centering
\includegraphics[scale=0.40]{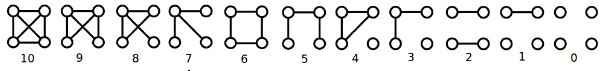}
\caption{Set of all non isomorphic graphlets of size $4$}
\label{graphlets}
\end{figure}

\begin{figure}[ht]
\centering
\includegraphics[scale=0.65]{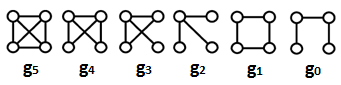}
\caption{Non isomorphic connected graphlets of size $4$}
\label{connectedgraphlets}
\end{figure}

Graphlets are fixed size non isomorphic induced subgraphs of a large graph. Typical graphlet sizes considered in applications are $3, 4$ and $5$.
For example, Figure \ref{graphlets} shows all possible non isomorphic size $4$ graphlets. There are $11$ of them of which $6$ are connected. We denote by $D_l$, the set of all size $l$ graphlets that are connected. The set $D_4$ is shown in Figure \ref{connectedgraphlets}.



\begin{mydef}[Graphlet Vector]
For a given $l$, the graphlet vector $f_G$ for a given graph $G$ is a frequency vector of dimension $|D_l|$ where its $i$th component corresponds to the number of occurrences of the $i$th graphlet of $D_l$ in $G$. Here, the graphlet vector $f_G$ is assumed to be normalized by the $L_2$ norm $||f_G||_2$. In \cite{shervashidze2009efficient}, the graphlet vector is normalized by the $L_1$ norm $||f_G||_1$. We use $L_2$ normalization instead as it is directionally invariant.
\end{mydef}
If graphs $G$ and $G'$ are isomorphic then clearly their corresponding graphlet vectors $f_G$ and $f_{G'}$ are identical.
But the reverse need not be true in general. But, it is conjectured that given two graphs $G$ and $G'$ of $n$ vertices and their corresponding graphlet vectors $f_G$ and $f_{G'}$ with respect to $n-1$ sized graphlets $D_{n-1}$, graph $G$ is isomorphic to $G'$ if $f_G$ is identical to $f_{G'}$ \cite{shervashidze2009efficient}. The conjecture has been verified for $n \le 11$ \cite{shervashidze2009efficient}. Kernels based on similarity of graphlet vectors provide a natural way to express similarity of underlying graphs.  
\begin{mydef}[Graphlet Kernel]
Given two graphs $G$ and $G'$, let $f_G$ and $f_{G'}$ be their corresponding graphlet frequency vectors with respect to size $l$ graphlets for some fixed $l$. The graphlet kernel value $K(G, G')$ is defined as the dot product of $f_G$ and  $f_{G'}$. That is, $K(G,G') = f^{T}_{G}f_{G'}$
\end{mydef}

Graphlet vectors are in fact an explicit embedding of graphs into a vector space whose dimension is $|D_l|$ if size $l$ graphlets are used. Graphlet kernels have been shown to give better classification accuracies in comparison to other graph kernels like random walk kernel and shortest path kernel for certain applications \cite{shervashidze2009efficient}. Values of $K(G, G') \in [0, 1]$ and larger values of $K(G, G')$ indicate higher similarity between $G$ and $G'$.


\begin{myprobstmt*}
Let $K(\cdot,\cdot)$ be a graphlet kernel based on size $l$ graphlets for some fixed $l$. Given a large connected graph $G$ of size $n$ and a connected query graph $Q$ of size $n_q$ with $n > n_q$, find a subset $V^*$ of vertices in $G$ such that its  induced subgraph $G^*$ in $G$ maximizes $K(Q, G^*)$.
\end{myprobstmt*}


 

\section{Graphlet vector based vertex labeling}

Computing vertex labels that capture topological neighborhood information of corresponding vertices in the graph and comparing vertex neighborhoods using their labels is crucial in our matching algorithm. Our vertex labels are graphlet vectors of their corresponding neighborhood subgraphs.

Given a fixed positive integer $t$ and graph $G$, let $N(v)$ denote the depth $t$ neighbors of vertex $v$ in $G$. That is, $N(v)$ is the subset of all vertices in $G$ (including $v$) that are reachable from $v$ in $t$ or less edges. Let $H_v$ denote the subgraph induced by vertices $N(v)$ in $G$. We denote by $f_v$, the graphlet vector corresponding to the graph $H_v$, with respect to size $l$ graphlets for some fixed $l$. 
We note that for defining the graphlet vector $f_v$ for a vertex, there are two implicit parameters $l$ and $t$. To avoid overloading the notation, we assume them to be some fixed constants and specify them explicitly when required. Values of $l$ and $t$ are parameters to our final algorithm.

For each vertex $v$ of the graph, its vertex label is given by the vector $f_v$. Given vertex labels $f_u$ and $f_v$ for vertices $u$ and $v$,
we denote by $s(u, v)$ the similarity between labels of $f_u$ and $f_v$, given by their dot product as

\begin{equation} \label{simscore}
s(u,v) = f^{T}_uf_{v}
\end{equation}

Values of $s(u, v) \in [0, 1]$ and larger values of $s(u, v)$ indicate higher topological similarity between neighborhoods of vertices $u$ and $v$. Computing the vertex labels of the target graph is done in the preprocessing phase. Implementation details of the vertex labeling algorithm are discussed in the next section. 

\section{Our Algorithm}

Our subgraph similarity search algorithm has two major phases: one time pre-processing phase and the query graph matching phase.
Each of these phases comprise sub-phases as given below. Details of each of these subphases is discussed in the subsequent sections.


\begin{enumerate}
\item[\textit{A.}] \textit{Pre-processing Phase: } This phase has two subphases:
\begin{enumerate}[1)]
\item In this phase, vertex labels $f_v$ of all the vertices of the target graph $G$ are computed.
\item k-d tree based nearest neighbor data structure on the vertices of $G$ using their label vectors $f_v$ is built.
\end{enumerate}
\item[\textit{B.}] \textit{Matching Phase: } This phase is further divided into four subphases:
\begin{enumerate}[1)]
\item Selection Phase: In this phase, vertex labels $f_v$ for vertices of the query graph $Q$ are computed first. Each vertex $u$ of the query graph then selects a subset of vertices from the target graph $G$ closest to $u$ based on their Euclidean distance.
\item Seed Match Generation Phase: In this phase, 
a one to one mapping of a subset of query graph vertices to target graph vertices is obtained with highest overall similarity score. Subgraph induced by the mapped vertices in the target graph is called the seed match.
The seed match is obtained by solving a maximum weighted bipartite matching problem.
\item Match Growing Phase: The above seed match is used as a basis to compute the final match for $Q$. 
\item Match Completion Phase: This phase tries to match those vertices in $Q$ that are still left unmatched in the previous phase.
\end{enumerate}
\end{enumerate}

\subsection{Pre-processing Phase}
\subsubsection{Computation of vertex labels $f_v$}
In this phase, vertex label $f_v$ for each vertex $v$ of the target graph $G$ is computed first.
To compute $f_v$, we require parameter values $t$ and $l$. These two values are assumed to be provided as parameters to the search algorithm. For each vertex $v$, a breadth first traversal of depth $t$ is performed starting from $v$ to obtain the depth $t$ neighborhood $N(v)$ of $v$. The graph $H_v$ induced by the vertex set $N(v)$ is then used to compute the graphlet vector $f_v$ as given in \cite{przulj2005supplementary}.
The pseudo code is given in Algorithm 1. 

Major time taken by the pre-processing phase is for computing the graphlet vector for $H_v$. In \cite{shervashidze2009efficient}, methods to improve its efficiency including sampling techniques are discussed. We do not make use of sampling technique in our implementation.  We remark that finding the graphlet frequencies can easily be scaled to parallel computing frameworks or distributed computing frameworks such as map-reduce.

\begin{algorithm} \label{algorithm1}
\caption{Compute label $f_v$ for vertex $v$}
\begin{algorithmic}[1]
\renewcommand{\algorithmicrequire}{\textbf{Input:}}
\renewcommand{\algorithmicensure}{\textbf{Output:}}
\REQUIRE Graph $G$, vertex $v$, BFS depth $t$, graphlet size $l$
\ENSURE  Label vector $f_v$
\STATE Run BFS on $G$ starting from $v$ till depth $t$. Let $N(v)$ be the set of vertices visited including $v$. 
\STATE Identify the induced subgraph $H_v$ of $G$ induced by $N(v)$.
\STATE Compute graphlet vector $f_v$ for graph $H_v$. 
\STATE Normalize $f_v$ by $||f_v||_2$.
\RETURN $f_v$ 
\end{algorithmic} 
\end{algorithm}

\subsubsection{Nearest neighbor data structure on $f_v$}
After computing vertex labels for $G$, a nearest neighbor data structure on the vertices of $G$ based on their label vectors $f_v$ is built. 
We use k-d trees for nearest neighbor data structure \cite{heineman2008algorithms}. k-d trees are known to be efficient when dimension of vectors is less than 20 \cite{heineman2008algorithms}. Since the typical graphlet size $l$ that we work with are $3, 4$ and $5$, the dimension of $f_v$ (which is $|D_l|$) does not exceed 10.

\subsection{Matching Phase}
In the following we describe the three subphases of matching phase.
\subsubsection{Selection Phase}
The vertex labels $f_v$ for all vertices of the query graph $Q$ are computed first using Algorithm 1. Let $R_v$ denote the set of $k$ vertices in $G$ that are closest to $v$ with respect to the Euclidean distance between their label vectors. In our experiments, we  usually fix $k$ as 10.
For each vertex $v$ of $Q$, we compute $R_v$ by querying the  k-d tree built in the pre-processing phase. Let $R$ denote the union of $R_v$ for each vertex $v$ of the $n_q$ vertices of  $Q$. For the subsequent seed match generation phase, we will only consider the vertex subset $R$ of $G$. Clearly size of $R$ is at most $k . n_q$ which is typically much smaller than the number of vertices in $G$.


\subsubsection{Seed Match Generation Phase} 
In this phase, we obtain a one to one mapping of a subset of vertices of the query graph $Q$ to the target graph $G$ with highest overall similarity score. We call the subgraph induced by the mapped vertices in $G$ as the seed match. To do this, we define a bipartite graph $(V(Q), R)$ with weighted edges, where one part is the vertex set $V(Q)$ of the query graph $Q$ and the other part is the pruned vertex set $R$ of $G$ obtained in the previous step. The edges of the bipartite graph and their weights are defined as follows. Each vertex $v$ in the part $V(Q)$ is connected to every vertex $w$ in $R_v \subseteq R$, where $R_v$ is the set of $k$ nearest neighbors of $v$ in $G$ as computed in the previous step. 

The weight $\lambda(v, w)$ for the edge $(v, w)$ is defined in the following manner.  
Let $0 < \alpha < 1$ be a fixed scale factor which is provided as a parameter to the search algorithm. We recall that vertex $v$ belongs to query graph $Q$ and vertex $w$ belongs to target graph $G$ and $s(v, w)$ given by equation (\ref{simscore}) denote the similarity between their label vectors $f_v$ and $f_w$.
Let $V_w$ denote the neighbors of vertex $w$ in graph $G$ including $w$. Let $Q'$ denote the subset of $V(Q)$ excluding $v$ such that each vertex in $Q'$ is connected to at least one vertex in $V_w$ in the bipartite graph $(V(Q), R)$. In particular, for each vertex $u \in Q'$, let $s(u)$ denote the maximum $s(u, z)$ value among all its neighbors $z$ in $V_w$ in the bipartite graph. Now the weight $\lambda(v, w)$ for the edge $(v, w)$ of the bipartite graph is given by

\begin{equation} \label{modedgeweight}
\lambda (v, w) = \frac{\left( s(v, w)^\alpha + \sum_{u \in Q'} s(u)^\alpha\right)^{1/\alpha}}{(|Q'| + 1)}
\end{equation}

We now solve maximum weighted bipartite matching  on this graph to obtain a one to one mapping between a subset of vertices of $Q$ and the vertices of $G$. Defining edge weights $\lambda(v, w)$ to edge $(v, w)$ in the bipartite graph in the above fashion not only takes into account the similarity value $s(v, w)$, but also the strength of similarity of neighbors of $w$ in $G$ to remaining vertices in the query graph $Q$. By assigning edge weights as above, we try to ensure that among two vertices in $G$ with equal similarity values to a vertex in $Q$, the vertex whose neighbors in $G$ also have high similarity to vertices in $Q$ is preferred over the other in the final maximum weighted bipartite matching solution. 

Let $M$ denote the solution obtained for the bipartite matching. 
Let $Q_M$ and $G_M$ respectively denote the subgraphs induced by the subset of matched vertices from graphs $Q$ and $G$ under the matching $M$. The connectivity of $Q_M$ and $G_M$ may differ. For instance, the number of connected components in $G_M$ and $Q_M$ could differ. Therefore, we do not include all the vertices of $G_M$ in the seed match. Instead, we use the largest connected component of $G_M$ as a seed solution. That is, let $S_G \subset V(G)$ denote the subset of vertices in $G_M$ corresponding to a maximum cardinality connected component. Let $S_Q$ denote their corresponding mapped vertices  in $Q_M$. We call $S_G$ as a seed match.
The pseudo code for seed match computation is given in Algorithm 2. 

\begin{algorithm}
\caption{Computing seed match $S_G$ in $G$ and its mapped vertices $S_Q$ in $Q$} 
\begin{algorithmic}[1]
\renewcommand{\algorithmicrequire}{\textbf{Input:}}
\renewcommand{\algorithmicensure}{\textbf{Output:}}
\REQUIRE Vertex sets $V(Q)$, $R$ and $R_v$ for each $v \in V(Q)$ and their labels $f_v$, parameter $\alpha$
\ENSURE $S_G$ and $S_Q$
\STATE Construct bipartite graph $(V(Q), R)$ with edge weights given by $\lambda(v, w)$.
\STATE Compute maximum weighted bipartite matching $M$ on $(V(Q), R)$
\STATE Let $Q_M$ and $G_M$ respectively denote the subgraphs induced by vertices from $Q$ and $G$ in the matching $M$.
\STATE Compute largest connected component in $G_M$. Let $S_G$ denote the vertices in that component. Let $S_Q$ denote its mapped vertices in $Q_M$ under the bipartite matching $M$.
\RETURN $S_G$ and $S_Q$
\end{algorithmic} 
\end{algorithm}

\subsubsection{Match Growing Phase} 
After computing the seed match $S_G$ in $G$ and its mapped vertices $S_Q$ in $Q$, we use this seed match as the basis to compute the final match. The final solution is computed in an incremental fashion starting with empty match. In each iteration, we include a new pair of vertices $(v, w)$ to the solution, where $v$ and $w$ belongs to $G$ and $Q$ respectively.  In order to do this,  we maintain a list of candidate pairs and in each iteration, we include a pair with maximum similarity value $s(v, w)$ to the final solution. We use a max heap to maintain the candidate list. The candidate list is initialized with the mapped pairs between $S_G$ and $S_Q$ as obtained in the previous phase. Thus, the heap is initialized by inserting each of these mapped pairs $(v, w)$ with corresponding weight $s(v, w)$.  

We recall that the mapped pairs obtained from previous phase have stronger similarity with respect to the modified weight function $\lambda(v, w)$. Higher value of $\lambda(v, w)$ indicates that not only $s(v, w)$ is high but also their neighbors share high $s()$ value. Hence they are more preferred in the solution over other pairs with similar $s()$ value. By initializing the candidate list with these preferred pairs, the matching algorithm tries to ensure that the incremental solution starts with these  pairs first and  other potential pairs are considered later. Also, because of the heap data structure, remaining pairs are considered in the decreasing order of their similarity value. Moreover, as will be discussed later, the incremental matching tries to ensure that the partial match in $G$ constructed so far is connected. For this, new pairs that are added to the candidate list are chosen from the neighborhood of the partial match between $G$ and $Q$.  

The incremental matching might still match vertex pairs with low $s()$ value if they are available in the candidate list. Candidate pairs with low $s()$ values should be treated separately as there could be genuine pairs with low $s()$ value.  For instance, consider boundary vertices of an optimal subgraph match in $G$. Boundary vertices are also connected to vertices outside the matched subgraph. Hence, their local neighborhood structure is different from their counterpart in the query graph. In other words, their corresponding graphlet vectors can be very dissimilar and their similarity value $s()$ can be very low even though they are expected to be matched in the final solution. In order to find such genuine pairs, we omit pairs with similarity value below some fixed threshold $h_1$ in this phase and such pairs are handled in the next phase. 

In each iteration of the  incremental matching, a pair $(v, w)$ with maximum $s(v, w)$ value is removed from the candidate heap and added to the final match.  After this, the candidate list is modified as follows. We recall that $v$ and $w$ belong to $G$ and $Q$ respectively. We call a vertex unmatched if it is not yet present in the final match. The algorithm maintains two invariants:  (a) the pairs present in the candidate list are one to one mappings and (b) a query vertex that enters the candidate list will stay in the candidate list (possibly with multiple changes to paired partner vertex) until it is included in the final match. Let  $U_v$ denote the unmatched neighbors of  $v$ in $G$ that are also not present in the candidate list.  Let $U_w$ denote the unmatched neighbors $w$ in $Q$. For each query vertex $y$ in $U_w$, let $x$ be a vertex in $U_v$ with maximum similarity value $s(x, y)$. We add $(x, y)$ to the candidate list if $y$ is absent in the list and $s(x,y) \ge h_1$. If $y$ is already present in the candidate list, then replace the current pair for $y$ with $(x, y)$ if $s(x, y)$ has a higher value.
The incremental algorithm is given in Algorithm 3. The candidate list modification is described in Algorithm 4.

\begin{algorithm}
\caption{Incremental Matching}
\begin{algorithmic}[1]
\renewcommand{\algorithmicrequire}{\textbf{Input:}}
\renewcommand{\algorithmicensure}{\textbf{Output:}}
\REQUIRE Seed match $S_G$ and its mapped vertices $S_Q$, threshold $h_1$
\ENSURE  Partial match $F$
\STATE Initialize $F$ to empty set.
\STATE Initialize the candidate list max heap with mapped pairs $(v, w)$ of the seed match where $s(v, w) \ge h_1$. 
\WHILE{candidate list is not empty}
	\STATE Extract maximum weight candidate match $(v, w)$
	\STATE Add  $(v, w)$ to $F$
	\STATE \textbf{updateCandidateList}(candidate list, $(v, w)$, $h_1$,$F$)
\ENDWHILE
\RETURN $F$
\end{algorithmic} 
\end{algorithm}

\begin{algorithm}
\caption{updateCandidateList}
\begin{algorithmic}[1]
\renewcommand{\algorithmicrequire}{\textbf{Input:}}
\renewcommand{\algorithmicensure}{\textbf{Output:}}
\REQUIRE candidate list, $(v, w)$, $h_1$ and $F$
\STATE Compute  $U_v$ which is the set of unmatched neighbors of $v$ in $G$ that are also not present in candidate list.
\STATE Compute  $U_w$ which is the set of unmatched neighbors of $w$ in $Q$.
\FORALL{vertex $y \in U_w$}
	\STATE Find $x \in U_v$ with maximum $s(x, y)$ value.
	\IF{$y$ does not exist in candidate list}
		\STATE Include $(x, y)$ in the candidate list if $s(x, y) \ge h_1$.
	\ELSE
		\STATE Replace existing pair for $y$ in the candidate list with $(x, y)$ if $s(x, y)$ has higher value.
	\ENDIF
\ENDFOR
\end{algorithmic} 
\end{algorithm}

\subsubsection{Match Completion Phase}

In this phase, vertices of the query graph $Q$ that are left unmatched in the previous phase due to similarity values below the threshold $h_1$ are handled. Typically, boundary vertices of the final matched subgraph in $G$ remain unmatched in the previous phase. As discussed earlier, this is because, such boundary vertices in $G$ and their matched partners in $Q$ have low $s()$ value as their local neighborhood topologies vastly differ.  Hence using neighborhood similarity for such pairs is ineffective. To handle them, we try to match unmatched query vertices with unmatched neighbors of the current match $F$ in $G$.  Since the similarity function $s()$ is ineffective here, we use a different similarity function to compare potential pairs.  Let $X$ denote the set of unmatched neighbors of the current match $F$ in $G$. Let $Y$ denote the set of unmatched query vertices.  Let $v \in X$ and let $w \in Y$. We define the similarity $c(v, w)$ as follows. Let $Z_v$ denote the matched neighbors of $v$ in target graph $G$ and let $Z_w$ denote the matched neighbors of $w$ in query graph $Q$. Let $Z'_v$ denote the matched partners of $Z_v$ in $Q$. We now define $c(v, w)$ using the standard Jaccard similarity coefficient as

\begin{equation}
c(v, w) = \frac{|Z'_v \cap Z_w|}{|Z'_v \cup Z_w|}
\end{equation}

We use a fixed threshold $h_2$ that is provided as parameter to the algorithm. We now define a bipartite graph $(X, Y)$ with edge weights as follows. For each $(v, w) \in X \times Y$, insert an edge $(v, w)$ with weight $c(v, w)$ in the bipartite graph if $c(v, w) \ge h_2$. Compute maximum weighted bipartite graph matching on this bipartite graph and include the matched pairs in the final solution $F$. In our experiments, size of $Y$ (number of unmatched query graph vertices) is very small. The pseudo code is given in Algorithm 5.

\begin{algorithm}
\caption{Match Completion}
\begin{algorithmic}[1]
\renewcommand{\algorithmicrequire}{\textbf{Input:}}
\renewcommand{\algorithmicensure}{\textbf{Output:}}
\REQUIRE Partial match $F$ and threshold $h_2$
\ENSURE  Final match $F$
\STATE Let $X$ denote the set of unmatched neighbors of the match $F$ in $G$.
\STATE Let $Y$ denote the set of unmatched vertices in $Q$.
\STATE Construct bipartite graph $(X, Y)$ by introducing all edges $(v, w)$ with edge weight $c(v, w)$ if $c(v, w) \ge h_2$.
\STATE Compute maximum weighted bipartite matching.
\STATE Add each of these matches to $F$
\RETURN $F$
\end{algorithmic} 
\end{algorithm}

We remark that our searching algorithm finds the matched subset of vertices in $G$ and also their corresponding mapped vertices in the query graph $Q$. 

\section{Experimental Results}

In this section, we conduct experiments on various real life graph data sets \cite{snapnets} including social networks, collaboration networks, road networks, youtube network, amazon network  and on  synthetic graph data sets. 

\subsection{Experimental Data sets}

\begin{enumerate}
\item[] \textbf{Social Networks}: We conduct experiments on facebook and google plus undirected graphs provided by Stanford Large Network Dataset Collection (SNAP) \cite{snapnets}. Facebook graph contains around 4K vertices and 88K edges. In this graph vertices represent anonymized users and an undirected edge connects two friends. google plus graph contains 107K vertices and 13M edges. google plus graph also represents users as vertices and an edge exists between two friends. The data set also contains list of user circles (user communities), where user circle is specified by its corresponding set of vertices. We use these user circles as query graphs and they are queried against the entire facebook network. We also query facebook circles against google plus network to find similar circles across networks. We also experiment querying facebook circles against facebook network after introducing random noise to the facebook network.
\\
\item[] \textbf{DBLP Collaboration Network}: We use the DBLP collaboration network  downloadable from \cite{DBLP}.
This network has around 317K vertices and 1M edges. The vertices of this graph are authors who publish in any conference or journal and an edge exists between any two co-authors. All the authors who contribute to a common conference or a journal form a community. The data set provides a list of such communities by specifying its corresponding  set of vertices. We use such communities as query graphs.
\\
\item[] \textbf{Youtube Network}: Youtube network is downloaded from \cite{snapnets}. Network has about 1M vertices and 2M edges. Vertices in this network represent users and an edge exists between two users who are friends. In youtube, users can create groups in which other users can join. The data set provides a list of user groups by specifying its corresponding set of vertices. We consider these user-defined groups as our query graphs.
\\
\item[] \textbf{Road Network}: We use the road network of California obtained from \cite{snapnets} in our experiments. This network has around 2M vertices and 3M edges. Vertices of this network are road endpoints or road intersections and the edges are the roads connecting these intersections. We use randomly chosen subgraphs from this network as query graphs.
\\
\item[] \textbf{Amazon Network}: Amazon network is a product co-purchasing network downloaded from \cite{snapnets}. This network has around 334K vertices and 925K edges. Each vertex represents a product and an edge exists between the products that are frequently co-purchased \cite{snapnets}. All the products under a certain category form a product community. The data set provides a list of product communities by specifying its corresponding set of vertices. We use product communities as query graphs and we query them against the amazon network.

\end{enumerate}

The statistics of the data sets used are listed in Table \ref{datastats}.

\begin{table}[ht]
	\caption{Data set Statistics}
	\centering
	\begin{tabular}{|c|c|c|}
		\hline
		\textbf{Data Set} & \textbf{\#vertices} &\textbf{\#edges}\\
		\hline
		Facebook & 4039 & 88234\\
		\hline
		Google Plus & 107614 & 13673453\\
		\hline
		DBLP & 317080 & 1049866\\
		\hline
		Amazon & 334863 & 925872\\
		\hline
		Youtube & 1134890 & 2987624\\
		\hline
		Road Network & 1965206 & 2766607\\
		\hline
	\end{tabular}
	\label{datastats}
\end{table}

\subsection{Experimental Setup}

All the experiments are carried out on a 32 core 2.60GHz Intel(R) Xeon(R) server with 32GB RAM. The server has Ubuntu 14.04 LTS.
Our implementation uses Java 7. 

The computationally most expensive part of our algorithm is the computation of vector labels for all vertices of a graph.
The preprocessing phase that computes label vectors for each vertex of the graph is multi-threaded and thus executes on all 32 cores. Similarly, in the matching phase, computing label vectors for all vertices of the query graph is also multi-threaded and uses all 32 cores. Remaining phases use only a single core.

\subsection{Results}

To evaluate the accuracy of the result obtained by our similarity search algorithm, we compute the graphlet kernel value $K(Q, G^*)$  between the query graph $Q$ and the subgraph $G^*$ of $G$ induced by the vertices $V^*$ of the final match $F$ in $G$. We use this value to show the similarity between the query graph and our obtained match and we refer to this value as \emph{similarity score} in our experiments. We recall that similarity score lies in the range $[0, 1]$ where $1$ indicates maximum similarity.

There are six parameters in our algorithm: (1) graphlet size $l$, (2) BFS depth $t$ for vertex label computation, (3) value of $k$ for the $k$ nearest neighbors from $k$-d tree, (4) value of $\alpha$ in the edge weight function $\lambda$ and (5) similarity thresholds $h_1$ for match growing phase  and $h_2$ for match completion phase. In all our experiments we fix graphlet size $l$ as $4$. We performed experiments with different values of $k, \alpha, h_1$ and $h_2$ on different data sets. Based on the results, we chose ranges for these parameters. The value of $k$ is chosen from the range $5$ to $10$. Even for million vertex graphs, $k=10$ showed good results. We fix scaling factor $\alpha$ to be $0.3$ and the thresholds $h_1$ and $h_2$ to be $0.4$ and $0.95$ respectively.

\textit{Experiment 1: } This experiment shows the effect of bfs depth $t$ on the final match. We performed experiments with different values of $t$. We observed that after the depth of 2, there is very little change in the similarity scores of the final match. But as the depth increases the time to compute graphlet vectors also increases. Thus, the bfs depth $t$ was taken to be 2 for most of our experiments. Table \ref{scvsd} shows the similarity scores of querying amazon communities on amazon network and and DBLP communities on DBLP collaboration network for different values of $t$. These results are averaged over 150 queries.

\begin{table}[ht]
	\caption{Experiment 1 : Similarity Score vs. $t$}
	\centering
	\begin{tabular}{|c|c|c|c|}
		\hline
		\textbf{Data Set} & \textbf{$t$=2} & \textbf{$t$=3} & \textbf{$t$=4}\\
		\hline
		Amazon & 0.9999823 & 0.9999851 & 0.9999858\\
		\hline
		DBLP & 0.9999942 & 0.9999896 & 0.9999917\\
		\hline
	\end{tabular}
	\label{scvsd}
\end{table}

\textit{Experiment 2: } For each of the data sets discussed earlier, we perform subgraph querying against the same network. For each network, we use the given communities as query graphs and measure the quality of the search result. That is, we query facebook communities against facebook network, DBLP communities against DBLP network, youtube groups against youtube network and amazon product communities against amazon network. For road network, we use randomly chosen induced subgraphs from the network as query graph. Second column of Table \ref{sc} shows the similarity score of the match. All the results are averages over 150 queries.  The average community (query graph) size is around 100 for facebook, around 40 for DBLP, around 50 for youtube and around 300 for amazon. Query graphs for road network have about 500 vertices.

To validate the quality of our solution,  we do the following for each of the network. We compute the similarity score between random induced subgraphs from the same network. These random subgraphs contain 100 vertices. We also compute the similarity score between different communities from the same network. All results are averaged over 150 scores. Table \ref{sc} shows the result. The results show that the similarity score  of our match close to 1 and is significantly better than scores between random subgraphs and scores between communities in the same network. For road network, the third column shows the average similarity between its query subgraphs.

\begin{table}[ht]
	\caption{Experiment 2 : Similarity Scores. Second column shows the average similarity score between query graph and the computed match. The query graphs are the given communities. Third column shows the average similarity score between random subgraphs. Fourth column shows average similarity score between communities }
	\centering
	\begin{tabular}{|c|c|c|c|}
		\hline
		\textbf{DataSet} & \textbf{Query graph \& } & \textbf{Between} & \textbf{Between}\\
		\textbf{} & \textbf{Final Match} & \textbf{ Random } & \textbf{Communities}\\
		\textbf{} & \textbf{} & \textbf{ Subgraphs} & \textbf{}\\
		\hline
		Facebook & 0.944231 & 0.702286 & 0.787296\\
		\hline
		DBLP & 0.975137 & 0.443763 & 0.6144779\\
		\hline
		Amazon & 0.999982 & 0.663301 & 0.624756\\
		\hline
		Youtube & 0.998054 & 0.311256 & 0.524779\\
		\hline
		Road Network& 0.899956 & 0.770492 & 0.599620\\
		\hline
	\end{tabular}
	\label{sc}
\end{table}

Table \ref{stats} shows the $\#exactMatches$ which is the number of queries that yielded the exact match out of the 150 queries (query graph is a subgraph of the network), and $\#inPruned$ - the percentage of queries where the vertices of the exact target match are present in the pruned subset of vertices $R$ of target graph $G$ obtained after the selection phase. 
Table \ref{stats} shows that, for about $30\%$ of the query graphs, our algorithm identifies the exact match. Also, for about $75\%$ of the queries, vertices of the ideal match are present in our pruned set of vertices $R$ in the target graph after selection  phase.

\begin{table}[ht]
	\caption{Experiment 2 : Exact Match Statistics}
	\centering
	\begin{tabular}{|c|c|c|}
		\hline
		\textbf{Data Set} & $\boldsymbol{\#exactMatches}$ & $\boldsymbol{\#inPruned}$\\
		\textbf{} & (out of 150) & (percentage)\\
		\hline
		Facebook & 53 & 83\\
		\hline
		DBLP & 47 & 82\\
		\hline
		Amazon & 60 & 72\\
		\hline
	\end{tabular}
	\label{stats}
\end{table}

Table \ref{times} shows the timing results corresponding to \textit{Experiment 2}. The timing information is only for the matching phase and it excludes the one time pre-processing phase. Here $\delta$ denotes time taken (in secs) to compute the label vectors for all vertices of the query graph and $\tau$ the time taken (in secs) for the entire matching phase (including $\delta$). We recall that the label vector computation is implemented as multithreaded on 32 cores and the remaining part is executed as a single thread. It can be seen that the label vector computation is the computationally expensive part and the remaining phases take much lesser time.

\begin{table}[ht]
	\caption{Experiment 2 : Timing Results}
	\centering
	\begin{tabular}{|c|c|c|}
		\hline
		\textbf{DataSet} & $\boldsymbol{\delta}$(in sec) & $\boldsymbol{\tau}$(in sec)\\
		\hline
		Facebook & 0.213596 & 0.253706\\
		\hline
		DBLP & 0.159492 & 0.777687\\
		\hline
		Amazon & 0.199767 & 0.781500\\
		\hline
		Youtube & 0.225131 & 0.989452\\
		\hline
		Road Network& 0.216644 & 1.437619\\
		\hline
	\end{tabular}
	\label{times}
\end{table}

\textit{Experiment 3: }In all previous experiments, query graphs were induced subgraphs of the target network. In this experiment, we evaluate the quality of our solution when the query graph is not necessarily an induced subgraph of the target graph. For this, we conduct two experiments. In the first experiment, we use facebook communities as query graphs and query them against google plus network. To validate the quality of our solution, we measure the similarity score of the query graph with a random induced subgraph in the target graph with same number of vertices. In the second experiment, we create a modified facebook network by randomly removing 5\% its original edges. We use this modified network as the target graph and query original facebook communities in this target graph. Here also, we validate the quality of our solution by measuring the similarity score for the query graph with a random induced subgraph of same number of vertices in the target graph. Table \ref{approxmatch} shows the results. Values shown for both experiments are averaged over 150 scores.  The results show that similarity score of our match is close to 1 and is significantly better than a random match.

\begin{table}[ht]
	\caption{Experiment 3 : Similarity Scores. Second column shows the similarity score between query graph and match. Third column shows the score between query graph and a random subgraph}
	\centering
	\begin{tabular}{|c|c|c|}
		\hline
		\textbf{DataSet} & \textbf{Final Match} & \textbf{Random Subgraph}\\
		\hline
		Google Plus & 0.912241 & 0.600442\\
		\hline
		Facebook with random noise & 0.933662 & 0.701198\\
		\hline
	\end{tabular}
	\label{approxmatch}
\end{table}

\textit{Experiment 4: } We use our matching algorithm to identify dense subgraphs in large networks. In particular, we search for dense subgraphs in DBLP and google plus networks. For this, we first generate dense random graphs using the standard $G(n,p)$ model with $n = 500$ and $p = 0.9$. We now use these random graphs as query graphs and query them against the DBLP and google plus networks. We use the standard definition of density $\rho$ of a graph $H = (V,E)$ as

\begin{equation} \label{density}
\rho = \frac{2|E|}{|V| * |V - 1|} \in [0,1]
\end{equation}

The average density of our random query graphs is $0.9$. We queried these dense random graphs against DBLP and google plus networks. Table \ref{densematch} shows the results. Column 2 shows the similarity score between query graph and obtained match. Column 3 shows the density $\rho$ for the obtained match. The results are averaged over 150 queries. Results show that the similarity score with matched result is close to 1 for google plus. For DBLP the score is close to 0.8 primarily because DBLP does not have dense subgraphs with about 500 vertices. Also, the density of the obtained match is close to that of the query graph, which is 0.9.

\begin{table}[ht]
	\caption{Experiment 4 : Dense Subgraph Match Results}
	\centering
	\begin{tabular}{|c|c|c|}
		\hline
		\textbf{DataSet} & \textbf{Similarity Score} & $\boldsymbol{\rho}$ for the match\\
		\hline
		Google Plus & 0.926670 & 0.812\\
		\hline
		DBLP & 0.799753 & 0.730\\
		\hline
	\end{tabular}
	\label{densematch}
\end{table}

\subsection{Scalability}

Computationally most expensive parts of our algorithm are the vertex label  computation for vertices of  query and target graphs. Since this is a one time preprocessing for the target graph, it can be easily scaled to a distributed framework using the standard map-reduce paradigm. Vertex label computation for each vertex can be a separate map/reduce job. Vertex label  computation for query graph is performed for every search. This  can also be parallelized using the standard OpenMP/MPI framework as each vertex label computation can be done in parallel.  As shown in the experimental results, remaining phases take much lesser time even with serial implementation. Parts of them can also be parallelized to further improve the search efficiency.

%
%
%


%

\bibliographystyle{IEEEtran}
\def\IEEEbibitemsep{0pt}
\bibliography{bare_conf}

\end{document}